\documentclass[aps,twocolumn,showpacs,preprintnumbers,floatfix]{revtex4}
\usepackage{graphicx}
\usepackage{epsfig}
\usepackage{dcolumn}
\usepackage{amsmath}

\def\be{\begin{equation}}
\def\ee{\end{equation}}
\def\bea{\begin{eqnarray}}
\def\eea{\end{eqnarray}}
\def\bsp{\be\begin{split}}

\def\bes{\be  \begin{split}}

\newcommand{\Rmnum}[1]{\expandafter\@slowromancap\romannumeral #1@}

\begin{document}

\title{Lattice Study of Radiative $J/\psi$ Decay to a Tensor Glueball}
\author{\small
Yi-Bo Yang,$^{1}$ Long-Cheng Gui,$^{1}$ Ying Chen,$^{1,}$\footnote{cheny@ihep.ac.cn} Chuan
Liu,$^{2}$ Yu-Bin Liu,$^{3}$ Jian-Ping Ma$^{4}$, and Jian-Bo~Zhang$^{5}$
\\(CLQCD Collaboration) } \affiliation{\small $^1$Institute of High Energy Physics and Theoretical
Center for Science
Facilities, Chinese Academy of Sciences, Beijing 100049, People's Republic of China \\
$^2$School of Physics and Center for High Energy Physics, Peking University, Beijing 100871, People's Republic of China\\
$^3$School of Physics, Nankai University, Tianjin 300071, People's Republic of China\\
$^4$Institute of Theoretical Physics, Chinese Academy of Sciences, Beijing 100190, People's Republic of China\\
$^5$Department of Physics, Zhejiang University, Zhejiang 310027, People's Republic of China }

\begin{abstract}
The radiative decay of $J/\psi$ into a pure gauge tensor glueball is studied in the quenched
lattice QCD formalism.  With two anisotropic lattices, the multipole amplitudes $E_1(0)$, $M_2(0)$
and $E_3(0)$ are obtained to be $0.114(12)(6)$ GeV, $-0.011(5)(1)$ GeV, and $0.023(8)(1)$ GeV,
respectively. The first error comes from the statistics, the $Q^2$ interpolation, and the continuum
extrapolation, while the second is due to the uncertainty of the scale parameter $r_0^{-1}=410(20)$
MeV. Thus, the partial decay width $\Gamma(J/\psi\rightarrow \gamma G_{2^{++}})$ is estimated to be
$1.01(22)(10)$ keV, which corresponds to a large branch ratio $1.1(2)(1)\times 10^{-2}$. The
phenomenological implication of this result is also discussed.
\end{abstract}

 \maketitle

Glueballs are exotic hadron states made up of gluons. Their existence is permitted by QCD but has
not yet been finally confirmed by experiment. In contrast to the scalar glueball, whose possible
candidate can be $f_0(1370)$, $f_0(1500)$, or $f_0(1710)$, the experimental evidence for the tensor
glueball is more obscure. Quenched lattice QCD studies predict the tensor glueball mass to be in
the range 2.2-2.4 GeV~\cite{prd56,prd60,prd73}, which is also supported by a recent $2+1$ flavor
full-QCD lattice simulation~\cite{Gregory:2012}. In this mass region, Mark
III~\cite{Baltrusaitis:1986} and BES~\cite{Bai:1996} have observed a narrow tensor meson
$\xi(2230)$ [now as $f_J(2220)$ in PDG\cite{PDG2012}] in the $J/\psi$ radiative decays with a large
production rate, whose features favor the interpretation of a tensor glueball. However, it was not
seen in the inclusive $\gamma$ spectrum~\cite{Kopke:1989} by the Crystal Ball Collaboration and in
$p\bar{p}$ annihilations to pseudoscalar
pairs~\cite{Barnes:1993,Hasan:1992,Hasan:1996,Bardin:1987,Sculli:1987,Evangelista:1997,Evangelista:1998,Buzzo:1997}.
So, the existence of $\xi(2230)$ [$f_J(2220)$] needs confirmation by new experiments, especially by
the BESIII experiment with the largest $J/\psi$ sample.
\par
It is well known that the production of glueballs is favored in $J/\psi$ decays because of the
gluon-rich environment. The radiative decay is of special importance, owing to its cleaner
background. So, the production rate of the tensor glueball in the decay can be an important
criterion for its identification. The decay has been studied only in a few theoretical
works~\cite{Li1981,Li:1987,Tenzo:1988,Melis:2004}. In these works, the tree-level perturbative QCD
approach is employed. Under certain assumptions, the helicity amplitudes of the decay are related
to the coupling of the two gluons to the tensor glueball. This coupling has been determined with
the quenched lattice QCD~\cite{prd73,Meyer2009}. Based on results of
Refs.~\cite{Li1981,Li:1987,Tenzo:1988} the branch ratio is estimated as $2\times
10^{-3}$~\cite{Li:2009}, but the theoretical uncertainties are not under control.
\par
In fact, the decay can be investigated directly from the numerical lattice QCD
studies~\cite{dudek06, Gui:2013}, which provide first principles calculations from the QCD
Lagrangian, especially in quenched lattice QCD. Quenched lattice QCD can be taken as a theory which
only consists of heavy quarks and gluons. In this theory amplitudes of the decay do not have an
absorptive part because of masses of states. Hence, the amplitudes can be directly calculated in
the theory in Euclidian space. It should be noted that it is still a challenging task for the
full-QCD lattice study of the decay because glueballs can be mixed with states of light quark
pairs. Nevertheless, the study of the decay in quenched QCD will give important information about
nonperturbative properties of glueballs.
\par
At the lowest order of QED, the amplitude for the radiative decay $J/\psi\rightarrow\gamma
G_{2^{++}}$ is given by
\begin{equation}
M_{r,r_\gamma,r_G}=\epsilon_{\mu}^*(\vec{q},r_\gamma)\langle
G(\vec{p}_f,r_G)|j^{\mu}(0)|J/\psi(\vec{p}_i,r)\rangle,
\end{equation}
where $\vec{q}=\vec{p}_i-\vec{p}_f$ is the momentum of the real photon, and $r$, $r_\gamma$, and
$r_G$ are the quantum numbers of the polarizations of $J/\psi$, the photon, and the tensor
glueball, respectively. $\epsilon_\mu(\vec{q},r_\gamma)$ is the polarization vector of the photon,
and $j^\mu$ is the electromagnetic current operator. The hadronic matrix element appearing in the
above equation can be obtained directly from a lattice QCD calculation of corresponding three-point
functions. On the other hand, these matrix elements can be expressed (in Minkowski space-time) in
terms of multipole form factors as follows:
\begin{eqnarray}\label{multipole}
&&\langle G(\vec{p}_f, r_G) | j^{\mu}(0) | J/\psi(\vec{p}_i,r)\rangle = \alpha_1^\mu
E_1(Q^2) + \alpha_2^{\mu}M_2(Q^2)\nonumber\\
&& + \alpha_3^\mu E_3(Q^2) + \alpha_4^\mu C_1(Q^2)+ \alpha_5^\mu C_2(Q^2)
\end{eqnarray}
where $\alpha_i^{\mu}$ are Lorentz-covariant kinematic functions of $p_i$ and $p_f$ (and specific
polarizations of the states), whose explicit expressions can be derived
exactly~\cite{Dudek2009,Yang:2012}, and $E_1(Q^2)$, $M_2(Q^2)$, $E_3(Q^2)$, $C_1(Q^2)$, and
$C_2(Q^2)$ are the form factors which depend only on $Q^2=-(p_i-p_f)^2$. Since $C_1(Q^2)$ and
$C_2(Q^2)$ vanish at $Q^2=0$, we focus on the extraction of the first three which are involved in
the calculation of the decay width $\Gamma(J/\psi\rightarrow \gamma G_{2^{++}})$ as
\begin{equation}\label{tensor_width}
\Gamma=\frac{4\alpha|\vec{p}_\gamma|}{27M_{J/\psi}^2}[|E_1(0)|^2+|M_2(0)|^2+|E_3(0)|^2],
\end{equation}
where $\alpha$ is the fine structure constant, and $\vec{p}_\gamma$ is the photon momentum with
$|\vec{p}_\gamma|=(M_{J/\psi}^2-M_G^2)/(2M_{J/\psi})$.

We use the tadpole-improved gauge action~\cite{prd56} to generate gauge configurations on
anisotropic lattices with the aspect ratio $\xi=a_s/a_t=5$, where $a_s$ and $a_t$ are the spatial
and temporal lattice spacings, respectively. Two lattices $L^3\times T=8^3\times 96(\beta=2.4)$ and
$12^3\times 144(\beta=2.8)$ are applied to check the effect of the finite lattice spacings. The
relevant input parameters are listed in Table~\ref{tab:lattice}, where $a_s$ values are determined
from $r_0^{-1}=410(20)$ MeV. Since glueball relevant study needs quite a large statistics, the
spatial extensions of both lattices are properly chosen to be $\sim 1.7$ fm according to the study
of the finite volume effect study of Ref.~\cite{prd73}, which is a compromise of the computational
resource requirement and negligible finite volume effects both for glueballs~\cite{prd73} and
charmonia. In the practice, we generated 5000 configurations for each lattice. The charm quark
propagators are calculated using the tadpole-improved clover action for anisotropic
lattices~\cite{chuan1,chuan2} with the bare charm quark masses set by the physical mass of
$J/\psi$, $M_{J/\psi}=3.097$ GeV, through which the spectrum of the $1S$ and $1P$ charmonia are
well reproduced~\cite{Yang:2012}. In practice, disconnected diagrams due to the charm and
quark-antiquark annihilation are expected to be unimportant according to the Okubo-Zweig-Iizuka
rule and therefore are neglected in the calculation of relevant two-point and three-point
functions.

\begin{table}[h]
\centering \caption{\label{tab:lattice} The input parameters for the calculation. Values for the
coupling $\beta$, anisotropy $\xi$, the lattice size, and the number of measurements are listed.
$a_s/r_0$ is determined by the static potential, and $a_s$ is estimated by $r_0^{-1}=410(20)$ MeV.}
\begin{ruledtabular}
\begin{tabular}{ccccccc}
 $\beta$ &  $\xi$  & $a_s/r_0$ &$a_s$(fm) & $La_s$(fm)&
 $L^3\times T$ & $N_{\rm conf}$ \\\hline
   2.4  & 5  & 0.461(4) & 0.222(2)(11) & $\sim 1.78$ &$8^3\times 96$ & 5000 \\
   2.8  & 5  & 0.288(2) & 0.138(1)(7) & $\sim 1.66 $&$12^3\times 144$ & 5000  \\
\end{tabular}
\end{ruledtabular}
\end{table}
\par
The calculations in this Letter are performed in the rest frame of the tensor glueball. One of the
key issues in our calculation is to construct optimal interpolating field operators which couple
dominantly to the pure gauge tensor glueball. This is realized by applying completely the same
scheme as that in the calculations of the glueball spectrum~\cite{prd60,prd73}. On the cubic
lattice, a tensor ($J=2$) state corresponds to the $E$ and $T_2$ irreducible representations of the
lattice symmetry group $O$. So, we build the $E$ and $T_2$ operators from a set of prototype Wilson
loops. By using different gauge-link smearing techniques, an operator set $\{\phi_\alpha^{(i)},
\alpha = 1,2,\ldots, 24\}$ of 24 different gluonic operators is constructed for each component of
the $T_2^{++}$ and $E^{++}$ representations, where the superscript $i$ labels the three components
of $T_2$ and two components of $E$. Finally, for each component, an optimal operator
$\Phi^{(i)}(t)=\sum v_{\alpha}\phi_\alpha^{(i)}(t)$ for the ground state tensor glueball is
obtained with the combinational coefficients $v_{\alpha}$ determined by solving the generalized
eigenvalue problem
\begin{equation}
\label{eigen} \tilde{C}^{(i)}(t_D){\bf v}^{(R)} = e^{-t_D\tilde{m}(t_D)}\tilde{C}^{(i)}(0){\bf
v}^{(R)},
\end{equation}
at $t_D=1$, where $\tilde{C}^{(i)}(t)$ is the correlation matrix of the operator set
\begin{equation}
\tilde{C}_{\alpha\beta}(t) =\frac{1}{N_t} \sum\limits_{\tau}\langle
0|{\phi}^{(i)}_\alpha(t+\tau){\phi}^{(i)}_\beta(\tau)|0\rangle.
\end{equation}
In addition, the glueball two-point functions are normalized as
\begin{eqnarray}
\label{glb_two}
 C^{i}(t)&=&\frac{1}{T}\sum\limits_{\tau}\langle
\Phi^{(i)}(t+\tau)\Phi^{(i)\dagger}(\tau)\rangle \nonumber\\
&\approx& \frac{|\langle 0|\Phi^{(i)}(0)|T_i\rangle|^2}{2M_TV_3}e^{-M_Tt}\approx e^{-M_Tt},
\end{eqnarray}
where $|T_i\rangle$ refers to $i$th component of the $T_2^{++}$ and $E^{++}$ glueball states. We
are assured that $C^i(t)$ can be well described by a single exponential $C(t)=We^{-M_Tt}$, with $W$
usually deviating from one by a few percents. It should be noted that the $SO(3)$ rotational
symmetry is broken on the lattice with a finite lattice spacing, and consequently the masses of
$T_2$ and $E$ glueballs are not necessarily the same, even though they converge to the same tensor
glueball mass in the continuum limit when the rotational invariance is restored. However, with the
two lattice spacings we used in this Letter, we observe that the difference of the two masses is
not distinguishable within errors, which implies that the effects of the rotational symmetry
breaking are not important. So, in the following, we neglect this symmetry breaking and assume that
the five components of the $T_2$ and $E$ and that of the corresponding spin-two state can be
connected by a normal transformation.
\par
We calculate the three-point functions in the rest frame of the tensor glueball with $J/\psi$
moving with a definite momentum $\vec{p}_f=2\pi \vec{n}/La_s$, where  $\vec{n}$ ranges from
$(0,0,0)$ to $(2,2,2)$. In order to increase the statistics additionally, for each configuration,
we calculate $T$ charm quark propagators $S_F(\vec{x},t;\vec{0},\tau)$ by setting a point source on
each time slice $\tau$, which permits us to average over the temporal direction when calculating
the three-point functions
\begin{eqnarray}
\Gamma_{i,\mu,j}^{(3)}(\vec{q};t_f,t) &=&
\frac{1}{T}\sum\limits_{\tau=0}^{T-1}\sum\limits_{\vec{y}} e^{-i\vec{q}\cdot \vec{y}} \langle
\Phi^{(i)}(t_f+\tau)\nonumber\\
&&\times J_\mu (\vec{y},t+\tau)O_{V,j}(\vec{0},\tau)\rangle\nonumber\\
&=&\sum\limits_{T,V,r}\frac{e^{-M_T (t_f-t)}e^{-E_V(\vec{q}) t}}{2M_T V_3 2E_V(\vec{q})}\nonumber\\
&&\times\langle 0|\Phi^{(i)}(0)|T_i\rangle \langle
T_i|J_\mu(0)|V(\vec{q},r)\rangle\nonumber\\
&&\times \langle V(\vec{q},r)|O_{V,j}^{\dagger}(0)|0\rangle,
\end{eqnarray}
where $J_\mu(x) =\bar{c}(x)\gamma_\mu c(x)$ is the vector current operator,
$O_{V,j}=\bar{c}\gamma_j c$ is the conventional interpolation field for $J/\psi$, and the summation
in the last equality is over all the possible states with different polarizations. In the rest
frame of the tensor glueball, the momentum of the initial $J/\psi$ is the same as that of the
current operator, say, $\vec{p}_V=\vec{q}$. The vector current $J_\mu (x)$, which is conserved in
the continuum limit, is no longer conserved on the lattice and requires a multiplicative
renormalization. The renormalization constant of spatial components of the vector current is
determined to be $Z_V^{(s)}=1.39(2)$ for $\beta=2.4$ and $Z_V^{(s)}=1.11(1)$ for
$\beta=2.8$~\cite{Gui:2013} using a nonperturbative scheme~\cite{dudek06}.
\par
The matrix elements $\langle T_i|J_\mu(0)|V(\vec{q},r)\rangle$ can be extracted from the above
three-point functions along with the two-point function of the glueball $C^i(t)$ and that of
$J/\psi$ $\Gamma_j^{(2)}(\vec{q},t)$,
\begin{equation}
\Gamma_j^{(2)}(\vec{q},t)=\sum\limits_{\vec{x}} e^{-i\vec{q}\cdot\vec{x}}\langle
0|O_{V,j}(\vec{x},t)O_{V,j}^\dagger(\vec{0},0)|0\rangle,
\end{equation}
which provide the information of $M_T$, $E_V(\vec{q})$, and the other two matrix elements.
According to Eq.~(\ref{glb_two}), one has approximately
\begin{equation}
\langle 0|\Phi^{(i)}(0)|T_i\rangle\approx \sqrt{2M_TV_3}.
\end{equation}

\begin{figure}
\includegraphics[height=4.5cm]{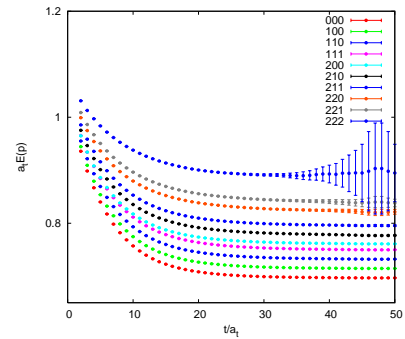}
\caption{The effective energy plot for $J/\psi$ with different spatial momenta. From top to bottom
are the plateaus for momentum modes, $\vec{p}=2\pi\vec{n}/L$, with $\vec{n}=(2,2,2)$, $(2,2,1)$,
$(2,2,0)$,
 $(2,1,1)$, $(2,1,0)$, $(2,0,0)$, $(1,1,1)$, $(1,1,0)$, $(1,0,0)$, and $(0,0,0)$. \label{psi_plat}}
\end{figure}
$M_T$ and $E_V(\vec{q})$ can be determined precisely from the two-point functions.
Figure~\ref{psi_plat} shows the nice effective energy plateaus of $J/\psi$ for typical momentum
modes at $\beta=2.4$. We also check the dispersion relation of $J/\psi$ and find the largest
deviation of squared speed of light $c^2$ from one is less than 4\%. The matrix elements $\langle
V(\vec{q},r)|O_{V,j}^{\dagger}(0)|0\rangle$ are included implicitly in the three-point and
two-point functions and can be canceled out by taking a ratio $R_{i,\mu,j}(\vec{q},t)$ for some
specific $\{i,\mu,j\}$ combinations
\begin{eqnarray}\label{eq_amp}
    R_{i,\mu,j}(\vec{q},t)&=&\Gamma^{(3)}_{i,\mu,j}(\vec{q},t_f,t)\times\nonumber\\
    &&\frac{\sqrt{4V_3 M_TE_V(\vec{q})}}{C^{i}(t_f-t)}\sqrt{\frac{\Gamma^{(2)}_j(\vec{q},t_f-t)}
    {\Gamma^{(2)}_j(\vec{q},t)\Gamma^{(2)}_j(\vec{q},t_f)}},\nonumber\\
\end{eqnarray}
which is expected to suppress the contamination from excited states of vector charmonia and should
be insensitive to the variation of $t$ in a time window. As such, the desired matrix elements can
be derived by the fit form
\begin{equation}
R_{i,\mu,j}(\vec{q},t)=\sum\limits_{r}\langle
T_i|J_{\mu}(0)|V(\vec{q},r)\rangle\epsilon_j(\vec{q},r)+\delta f(t),
\end{equation}
where $\vec{\epsilon}(\vec{q},r)$ is the polarization vector of $J/\psi$ and $\delta f(t)=a
e^{-mt}$ accounts for the residual contamination from excited states.
\par
\begin{figure}[t!]
\includegraphics[height=4cm]{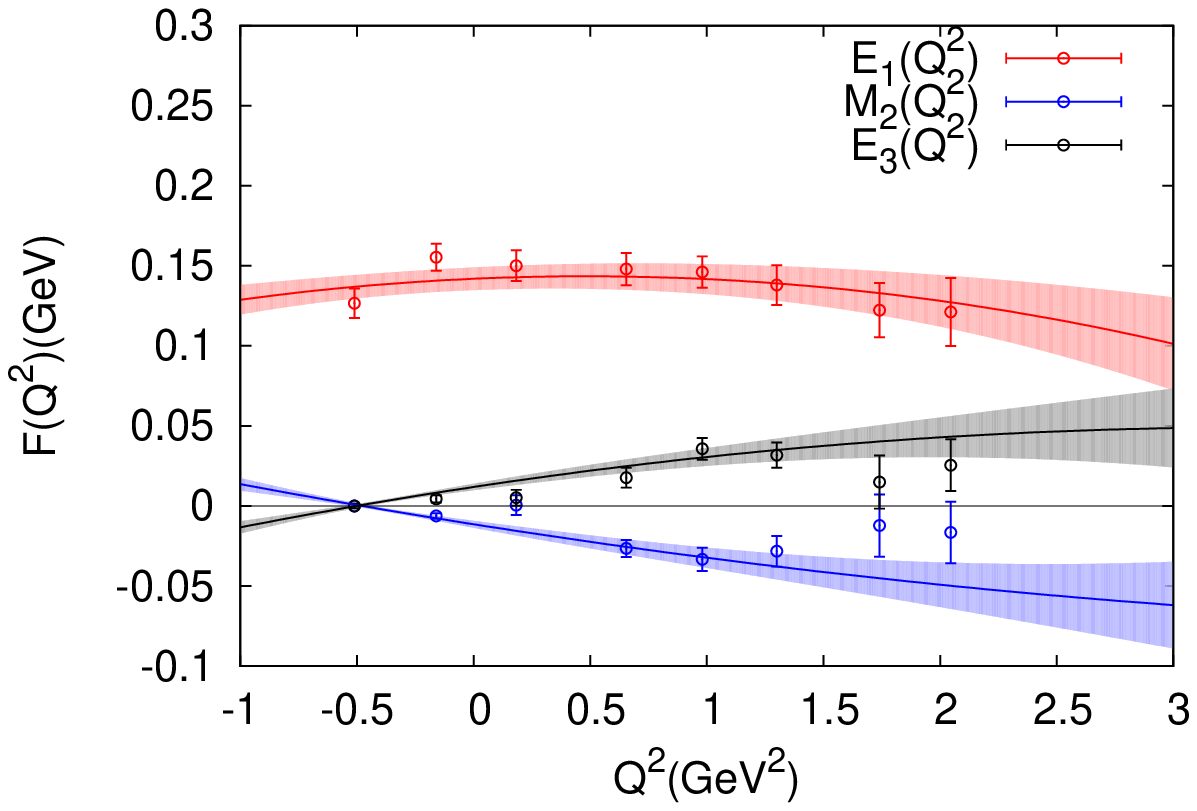}
\includegraphics[height=4cm]{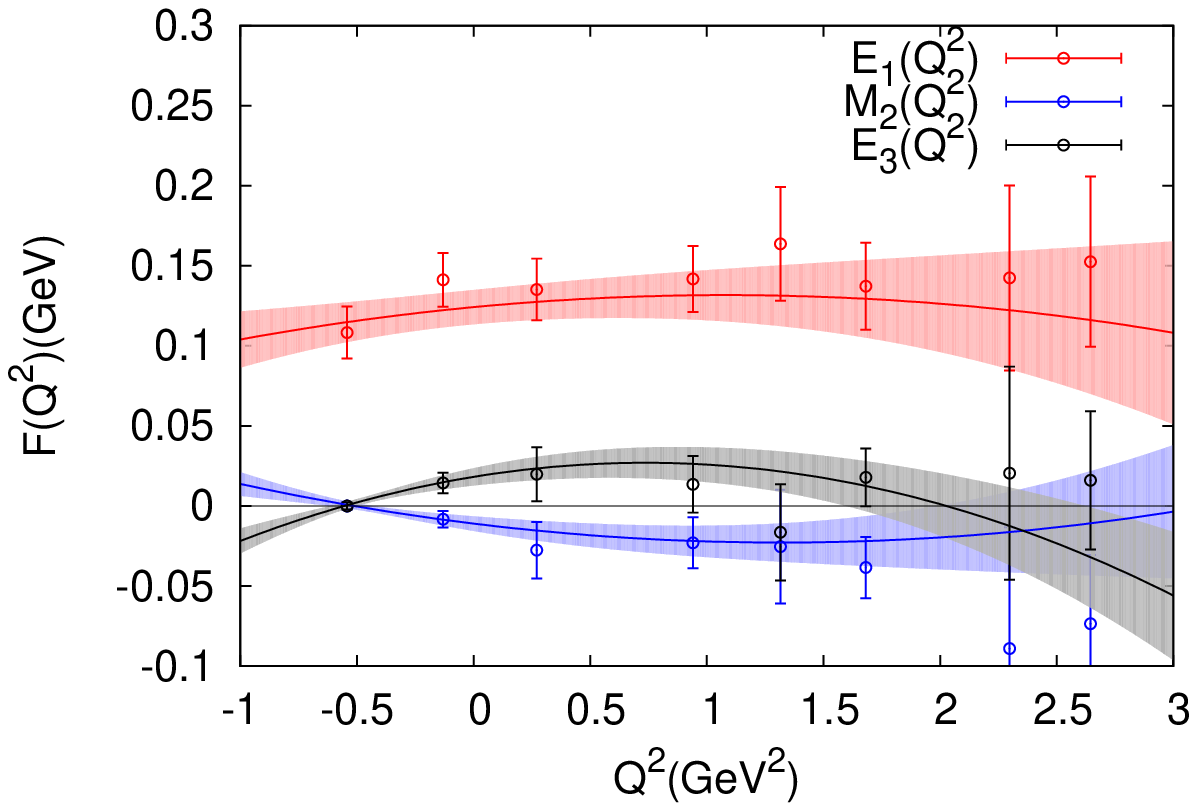}
\caption{\label{form}The extracted form factors $E_1(Q^2)$, $M_2(Q^2)$, and $E_3(Q^2)$ in the
physical units. The upper panel is for $\beta=2.4$ and the lower one for $\beta=2.8$. The curves
with error bands show the polynomial fit with $F_i(Q^2)=F_i(0)+a_iQ^2+b_iQ^4$.}
\end{figure}

In the data analysis, the 5000 configurations are divided into 100 bins and the average of 50
measurements in each bin is taken as an independent measurement. For the resultant 100
measurements, the one-eliminating jackknife method is used to perform the fit for the matrix
elements ($M_T$ and $E_V$ determined from two-point functions are used as known parameters).
Generally speaking, the time separations $t$ and $t_f-t$ should be kept large for the saturation of
the ground state, but we have to fix $t_f-t=1$ because of the rapid damping of the glueball signal
with respect to the noise. Fortunately this is justified to some extent by the optimal glueball
operators, which couple almost exclusively to the ground state. The second step of the data
analysis is to extract the form factors $E_1(Q^2)$, $M_2(Q^2)$, and $E_3(Q^2)$ at different
$Q^2=2E_V(\vec{q})M_T-M_V^2-M_T^2$ according to Eq.~(\ref{multipole}). Since the matrix elements
are measured from the same configuration ensemble, we carry out a correlated data fitting to get
these three form factors simultaneously with a covariance matrix constructed from the jackknife
ensemble described above. The symmetric combinations of the indices $\{i,\mu,j\}$ and the momentum
$\vec{q}$ which gives the same $Q^2$ are averaged to increase the statistics. In order to get the
form factor at $Q^2=0$, we carry out a correlated polynomial fit to the three form factors from
$Q^2=-0.5$ to 2.7 ${\rm GeV}^2$,
\begin{equation}\label{inter}
F_i(Q^2)=F_i(0)+a_iQ^2+b_iQ^4,
\end{equation}
where $F_i$ refers to $E_1$, $M_2$ or $E_3$.  Figure~\ref{form} shows the results of $F_i(Q^2)$ for
$\beta=2.4$ (upper panel) and $\beta=2.8$ (lower panel), where the data points are the calculated
values with jackknife errors, and the curves are the polynomial fits with jackknife error bands.
The corresponding interpolated $F_i(0)$'s are listed in Table~\ref{twobeta}. Note that the
renormalization constant $Z_V^{(s)}$ of the spatial components of the vector current is applied to
the final numerical values. We also fit the form factors by functions either linear in $Q^2$ in the
range $-0.5\,{\rm GeV^2}<Q^2<1.0\,{\rm GeV^2}$, or by adding a $Q^6$ term to Eq.~(\ref{inter}) in
the range $-0.5\,{\rm GeV^2}<Q^2<2.7\,{\rm GeV^2}$. The resultant $F_i(0)$'s are consistent with
that of Eq.~(\ref{inter}) within errors.

\par
The last step is the continuum extrapolation using the two lattice systems. After performing a
linear extrapolation in $a_s^2$, the continuum limits of the three form factors are determined to
be $E_1(0)=0.114(12)$ GeV, $M_2(0)=-0.011(5)$ GeV, and $E_3(0)=0.023(8)$ GeV, respectively.
Considering the uncertainty of the scale parameter $r_0^{-1}=410(20)$ MeV, which also introduces
$5\%$ error, the final result of the form factors is
\begin{eqnarray}
E_1(0)&=&0.114(12)(6)\,{\rm GeV}\nonumber\\
M_2(0)&=&-0.011(5)(1)\,{\rm GeV}\nonumber\\
E_3(0)&=&0.023(8)(1)\,{\rm GeV}.
\end{eqnarray}
Note that there is a pattern $|E_1(0)|\gg|M_2(0)|\sim |E_3(0)|$; hence, the decay width is
dominated by the value of $E_1(0)$. For the continuum value of the tensor glueball mass, we get
$M_G=2.37(3)(12)$ GeV (the second error is due to the uncertainty of $r_0$), which is compatible
with that in Ref.~\cite{prd73}. Thus, according to Eq.~(\ref{tensor_width}), we finally get the
decay width $\Gamma(J/\psi\rightarrow \gamma G_{2^{++}})=1.01(22)(10)$ keV. With the total width of
$J/\psi$, $\Gamma_{\rm tot}=92.9(2.8)$ keV~\cite{PDG2012}, the corresponding branching ratio is
\begin{equation}
\Gamma(J/\psi\rightarrow \gamma G_{2^{++}})/\Gamma_{\rm tot}=1.1(2)(1)\times 10^{-2}.
\end{equation}
\begin{table}[t]
\centering \caption{\label{twobeta} The tensor glueball masses $M_T$ as well as the form factors
$E_1(0)$, $M_2(0)$ and $E_3(0)$ for the two lattices with $\beta=2.4$ and 2.8. The last row gives
the continuum extrapolation. The uncertainty of the scale parameter $r_0$ has not been incorporated
yet.}
\begin{ruledtabular}
\begin{tabular}{ccccc}
 $\beta$  &  $M_T$(GeV)  &  $E_1$ (GeV) & $M_2$ (GeV) &  $E_3$ (GeV) \\
 \hline
   2.4   &  2.360(20)     &    0.142(07)       &  -0.012(2)  & 0.012(2) \\
   2.8   &  2.367(25)     &    0.125(10)       &  -0.011(4)  & 0.019(6) \\
   $\infty$ &  2.372(28)  &    0.114(12)       & -0.011(5)   & 0.023(8) \\
\end{tabular}
\end{ruledtabular}
\end{table}
The determined branching ratio is rather large. We admit that the calculation is carried out in the
quenched approximation, whose systematical uncertainty cannot be estimated easily without
unquenched calculations. A recent full-QCD lattice study of the mass spectrum of glueballs in
Ref.~\cite{Gregory:2012} indicates that there is no substantial correction of the masses of the
scalar and tensor glueball. Based on this fact, if the form factors also show similar behavior as
the masses, the unquenching effects might not change our result drastically. Of course, a full-QCD
lattice calculation would be very much welcome.

In experiments, the narrow state $f_J(2220)$ observed by Mark III and BES in the $J/\psi$ decay was
once interpreted as a candidate for the tensor glueball. But, the analysis of the processes
$p\bar{p}\rightarrow \pi\pi (K\bar{K})$ yields no indication of the narrow $f_J(2220)$ and sets an
upper bound for the branch ratios ${\rm Br}(f_J\rightarrow p\bar{p}){\rm Br}(f_J\rightarrow \pi\pi,
K\bar{K})$ (see the review article Ref.~\cite{Crede:2009} and the references therein). Combining
this with the results of Mark III and BES, a lower bound for the branching ratio is obtained to be
${\rm Br}[J/\psi\rightarrow \gamma f_J(2220)]>2.5\times 10^{-3}$~\cite{PDG2012}, which seems
compatible with our result. However, BESII with substantially more statistics does not find the
evidence of a narrow structure around 2.2 GeV of the $\pi\pi$ invariant mass spectrum in the
processes $J/\psi\rightarrow \gamma \pi\pi$~\cite{Ablikim06}, and {\it BABAR} does not observe it
in $J/\psi \rightarrow \gamma (K^+K^-, K_S K_S)$~\cite{Sanchez:2010}. Recently, based on 225
million $J/\psi$ events, the BESIII Collaboration performs a partial wave analysis of
$J/\psi\rightarrow \gamma\eta\eta$ and also finds no evident narrow peak for $f_J(2220)$ in the
$\eta\eta$ mass spectrum~\cite{Ablikim:2012}. So the existence of $f_J(2220)$ is still very weak.
Another possibility also exists that the tensor glueball is a broad resonance and readily decays to
light hadrons. Our result motivates a serious joint analysis of the radiative $J/\psi$ decay into
tensor objects in $VV$, $PP$, $p\bar{p}$, and $4\pi$ final states (where $V$ and $P$ stand for
vector and pseudoscalar mesons, respectively), among which $VV$ channels may be of special
importance since they are kinematically favored in the decay of a tensor meson.
\par
To summarize, we have carried out the first lattice study on the $E_1$, $M_2$, and $E_3$ multipole
amplitudes for $J/\psi$ radiatively decaying into the pure gauge tensor glueball $G_{2^{++}}$ in
the quenched approximation. With two different lattice spacings, the amplitudes are extrapolated to
their continuum limits. The partial decay width and branch ratio for $J/\psi\rightarrow \gamma
G_{2^{++}}$ are predicted to be $1.01(22)(10)$ keV and $1.1(2)(1)\times 10^{-2}$, respectively,
which imply that the tensor glueball can be copiously produced in the $J/\psi$ radiative decays if
it does exist. To date, the existence of $f_J(2220)$ needs confirmation and a broad tensor glueball
is also possible. Hopefully, the BESIII data will be able to clarify the situation.

\par
This work is supported in part by the National Science Foundation of China (NSFC) under Grants
No.10835002, No.11075167, No.11021092, No.11275169, and No.10975076. Y. C. and C. L. also
acknowledge the support of the NSFC and DFG (CRC110).

\end{document}